\documentstyle[12pt]{article}

\newcommand \beq{\begin{eqnarray}}
\newcommand \eeq{\end{eqnarray}}
\begin{document}
\title{DETERMINATION OF THE REACTION PLANE IN ULTRARELATIVISTIC NUCLEAR
COLLISIONS}
\author{Jean-Yves OLLITRAULT\footnotemark\\
Service de Physique Th\'eorique, CE-Saclay\\
91191 Gif-sur-Yvette, France\\ }
\maketitle
\footnotetext{$^*$ Member of CNRS.}
\begin{abstract}
If the particles produced in a nuclear collision undergo collective flow,
the reaction plane can in principle be determined through
a global event analysis. We show here that collective flow can be
identified by evaluating the reaction plane independently
in two separate rapidity intervals, and studying the correlation between
the two results. We give an analytical expression for the correlation
function between the two planes as a function of their relative angle.
We also discuss how this correlation function is related to the anisotropy
of the transverse momentum distribution.
\end{abstract}
\newpage
\def\square{\hbox{{$\sqcup$}\llap{$\sqcap$}}}   
\def\grad{\nabla}                               
\def\del{\partial}                              

\def\frac#1#2{{#1 \over #2}}
\def\smallfrac#1#2{{\scriptstyle {#1 \over #2}}}
\def\half{\ifinner {\scriptstyle {1 \over 2}}
   \else {1 \over 2} \fi}

\def\bra#1{\langle#1\vert}              
\def\ket#1{\vert#1\rangle}              

\def\simge{\mathrel{%
   \rlap{\raise 0.511ex \hbox{$>$}}{\lower 0.511ex \hbox{$\sim$}}}}
\def\simle{\mathrel{
   \rlap{\raise 0.511ex \hbox{$<$}}{\lower 0.511ex \hbox{$\sim$}}}}


\def\parenbar#1{{\null\!                        
   \mathop#1\limits^{\hbox{\fiverm (--)}}       
   \!\null}}                                    
\def\nunubar{\parenbar{\nu}}
\def\ppbar{\parenbar{p}}


\def\buildchar#1#2#3{{\null\!                   
   \mathop#1\limits^{#2}_{#3}                   
   \!\null}}                                    
\def\overcirc#1{\buildchar{#1}{\circ}{}}


\def\slashchar#1{\setbox0=\hbox{$#1$}           
   \dimen0=\wd0                                 
   \setbox1=\hbox{/} \dimen1=\wd1               
   \ifdim\dimen0>\dimen1                        
      \rlap{\hbox to \dimen0{\hfil/\hfil}}      
      #1                                        
   \else                                        
      \rlap{\hbox to \dimen1{\hfil$#1$\hfil}}   
      /                                         
   \fi}                                         %


\def\subrightarrow#1{
  \setbox0=\hbox{
    $\displaystyle\mathop{}
    \limits_{#1}$}
  \dimen0=\wd0
  \advance \dimen0 by .5em
  \mathrel{
    \mathop{\hbox to \dimen0{\rightarrowfill}}
       \limits_{#1}}}                           

\def\real{\mathop{\rm Re}\nolimits}     
\def\imag{\mathop{\rm Im}\nolimits}     

\def\tr{\mathop{\rm tr}\nolimits}       
\def\Tr{\mathop{\rm Tr}\nolimits}       
\def\Det{\mathop{\rm Det}\nolimits}     

\def\mod{\mathop{\rm mod}\nolimits}     
\def\wrt{\mathop{\rm wrt}\nolimits}     


\def\TeV{{\rm TeV}}                     
\def\GeV{{\rm GeV}}                     
\def\MeV{{\rm MeV}}                     
\def\KeV{{\rm KeV}}                     
\def\eV{{\rm eV}}                       

\def\mb{{\rm mb}}                       
\def\mub{\hbox{$\mu$b}}                 
\def\nb{{\rm nb}}                       
\def\pb{{\rm pb}}                       

%
\def\journal#1#2#3#4{\ {#1}{\bf #2} ({#3})\  {#4}}

\def\AdvPhys{\journal{Adv.\ Phys.}}
\def\AnnPhys{\journal{Ann.\ Phys.}}
\def\EurophysLett{\journal{Europhys.\ Lett.}}
\def\JApplPhys{\journal{J.\ Appl.\ Phys.}}
\def\JMathPhys{\journal{J.\ Math.\ Phys.}}
\def\LettNuovoCimento{\journal{Lett.\ Nuovo Cimento}}
\def\Nature{\journal{Nature}}
\def\NPA{\journal{Nucl.\ Phys.\ {\bf A}}}
\def\NPB{\journal{Nucl.\ Phys.\ {\bf B}}}
\def\NuovoCimento{\journal{Nuovo Cimento}}
\def\Physica{\journal{Physica}}
\def\PLA{\journal{Phys.\ Lett.\ {\bf A}}}
\def\PLB{\journal{Phys.\ Lett.\ {\bf B}}}
\def\PhysRev{\journal{Phys.\ Rev.}}
\def\PRC{\journal{Phys.\ Rev.\ {\bf C}}}
\def\PRD{\journal{Phys.\ Rev.\ {\bf D}}}
\def\PRL{\journal{Phys.\ Rev.\ Lett.}}
\def\PhysRept{\journal{Phys.\ Repts.}}
\def\ProcNatlAcadSci{\journal{Proc.\ Natl.\ Acad.\ Sci.}}
\def\ProcRoySoc{\journal{Proc.\ Roy.\ Soc.\ London Ser.\ A}}
\def\RevModPhys{\journal{Rev.\ Mod.\ Phys.}}
\def\Science{\journal{Science}}
\def\SovPhysJETP{\journal{Sov.\ Phys.\ JETP}}
\def\SovPhysJETPLett{\journal{Sov.\ Phys.\ JETP Lett.}}
\def\SovJNuclPhys{\journal{Sov.\ J.\ Nucl.\ Phys.}}
\def\SovPhysDoklady{\journal{Sov.\ Phys.\ Doklady}}
\def\ZPhys{\journal{Z.\ Phys.}}
\def\ZPhysA{\journal{Z.\ Phys.\ A}}
\def\ZPhysB{\journal{Z.\ Phys.\ B}}
\def\ZPhysC{\journal{Z.\ Phys.\ C}}
\setcounter{equation}{0}
\section{Introduction}

In searching for evidence of the formation of quark--gluon plasma in
ultrarelativistic nucleus--nucleus collisions,
one is led to address the question whether
the matter produced in these collisions can be considered,
at least locally, to be in thermal equilibrium. Local equilibrium
implies that the matter behaves collectively,
which may have observable consequences.
When collective flow is present, the evolution of the system
is determined by the pressure gradient, and is therefore strongly
influenced by geometrical factors (nuclear size, impact parameter...).
Effects of geometry on global event shapes have
made possible the detection and the study of collective flow of nuclear matter
at colliding energies up to 1~GeV per nucleon\cite{bevalac}.
However, few such studies have been undertaken at
ultrarelativistic energies, and no conclusion has been
drawn so far\cite{wa80}.

Most analyses at intermediate energies have been concerned
with the determination of the flow direction, which is the
direction of maximum kinetic energy flow.
However, the flow angle (angle between collision axis
and flow direction) decreases with increasing energy and
cannot be measured at ultrarelativistic energies;
this is because longitudinal momenta are much larger than
transverse momenta. When it can be measured,
the flow direction gives an experimental determination of the reaction plane,
which is the plane spanned by the collision axis and the impact parameter. The
latter can also be determined independently by measuring the transverse
momentum
transfer between target and projectile regions\cite{odyniec}.
Although this method has not been successful at SPS\cite{wa80}, it
has been recently argued that it should give results with
heavy nuclei\cite{amelin}.

Since the flow direction merges with the collision axis at
ultrarelativistic energies, whether or not there is collective behavior,
evidence for collective flow should rather be sought for in
the transverse directions.
Note further that while analyses carried out at intermediate energies
are mostly concerned with nucleons, the large number of mesons created
at ultrarelativistic energies in the central rapidity region
offers an opportunity to study collective flow, not only of
spectators and/or participant nucleons, but also among the produced particles.
Starting from these observations,
we have proposed a new signature in a previous work\cite{olli92}.
The idea is that for peripheral collisions,
the region where nucleon--nucleon collisions take place, when
projected onto the transverse plane, is anisotropic: it has
a smaller size in the direction of impact parameter than
in the perpendicular direction.
This causes the matter produced in the central rapidity
region  to flow preferentially in the
direction of the impact parameter, which results in a corresponding
anisotropy of the transverse momentum distribution.
This anisotropy should increase with impact parameter.

We propose to study collective
flow in the plane orthogonal to the principal flow direction (here, the
collision axis). This has also been
done  at lower energies\cite{squeeze}, where it was found that
matter escapes preferentially in the direction orthogonal to the reaction
plane. This was referred to as the {\it squeeze--out\/} effect.
By contrast, we predict a larger flow {\it in\/} the reaction plane.
However, this is only an apparent contradiction since the
effects that come into play are very different\cite{calcutta}.
Squeeze--out results
from an interaction between the participant nucleons,
which try to escape the fireball, and the spectators which block their
path in the reaction plane. But at ultrarelativistic energies,
the time it takes for the nuclei to cross each other is so short
that the particles produced in the central rapidity region do not
``see'' the spectators. Anisotropy
results from the interaction of particles in the central
rapidity region among themselves.

The main problem one usually encounters in global analyses is
that fluctuations occur due to the finite multiplicity, which
may hide collective effects\cite{gyulassy}. In this article we
would like to show how statistical and
dynamical effects can be separated in the anisotropy analysis.
The idea is that in a given event, one can perform two measurements
of the reaction plane by doing global analysis with two separate
subsets of the emitted particles (by selecting for instance the
particles produced in two separate rapidity intervals).
Then the difference between the two results gives a direct
measure of the dispersion due to statistical fluctuations.

In section 2 we recall the definitions of the variables we use in the global
analysis\cite{olli92}. In section 3 we give a general discussion of
finite multiplicity fluctuations. The main results are contained
in section 4 where we calculate, under quite general assumptions,
the correlation function between the two measurements of the reaction plane.
We show how to measure the ratio of dynamical to statistical
effects directly from the experiment. In section 5 we show how this
can be used to isolate the dynamical part in the anisotropy.
The relevance of this work to current and future experiments is discussed in
section 6.

\section{Global analysis}

In flow analyses, an event is characterized by means of global
observables describing its shape.
At ultrarelativistic energies,
we are only interested in the transverse directions.
We therefore define the $2\times 2$ transverse sphericity\cite{sphericity}
tensor $S^\bot$ by
\beq\label{sbot}
S^\bot_{ij}=\sum_{\nu=1}^{M}{w(\nu) u_i(\nu) u_j(\nu)}
\eeq
where $(u_1(\nu),u_2(\nu))$ is the unit vector parallel to the
transverse momentum of the $\nu^{\rm th}$ particle,
$w(\nu)$ is a weight, and the sum runs over all the
particles detected in a given rapidity interval.
If there is no particle
identification, $w(\nu)$ can be chosen to be equal to some
function of the transverse energy $E_T(\nu)$ deposited by the particle
in the calorimeter, for instance $w(\nu)=E_T^\beta (\nu)$, with
$\beta$ equal to some real constant. Collective flow
usually results not only in a larger number of particles
emitted in the flow direction, but also in a higher
energy per particle in this direction, so that $\beta$
should be taken positive, for instance $\beta=1$. Which is the best weight
to consider should be determined on the basis of a more
careful analysis. However, the results which are presented here do
not depend on such details.

$S^\bot_{ij}$ has three independent components and is therefore fully
determined by its two eigenvalues $f_1$ and $f_2$ (we choose $f_1\ge f_2$)
and the angle $\theta$ between the $x$--axis and the eigenvector associated
with $f_1$, with $-\pi/2\le\theta\le\pi/2$. Instead of $f_1$ and $f_2$ we
choose
the variables ${\cal E}$ and $\alpha$ defined by
\beq
{\cal E}&=&\tr {S^\bot}=f_1+f_2\nonumber\\
\alpha&=&{f_1-f_2\over f_1+f_2}
\eeq
$S^\bot$ can then be expressed as a function of ${\cal E}$, $\alpha$ and
$\theta$~:
\beq
S^\bot ={{\cal E}\over 2}\left(\matrix{
1+\alpha\,{\rm cos}\, 2\theta &   \alpha\,{\rm sin}\, 2\theta \cr
  \alpha\,{\rm sin}\, 2\theta & 1-\alpha\,{\rm cos}\, 2\theta}\right)
\eeq
$\alpha$ measures the relative difference between the eigenvalues of $S^\bot$,
i.e. the anisotropy of the momentum distribution.
As we said in the introduction,
collective flow results in anisotropy for peripheral
collisions\cite{olli92}.
Hydrodynamical calculations predict that $\alpha$ decreases
linearly with the multiplicity or the transverse energy (which
are measures of the impact parameter). The highest value of
$\alpha$, obtained for very peripheral collisions, is
about 0.25 to 0.3 for a Pb--Pb collision, and slightly less,
about 0.2, for a S--W collision.

There is a larger flow in the direction of the impact
parameter than in the direction perpendicular to the reaction
plane\cite{squeeze}.
Thus we expect that the principal axis associated with the larger eigenvalue
$f_1$
is the direction of the impact parameter. Since $S^\bot$ is directly
measurable,
this in turn gives an experimental measurement of the orientation of the
reaction plane. However, this estimate is reliable
only if the measured anisotropy originates from collective flow. Statistical
fluctuations related to the finite multiplicity also generate anisotropy, which
must be disentangled from the dynamical anisotropy created by collective
behavior. It is the purpose of this article to show how this can be
done experimentally.

\section{Finite multiplicity fluctuations}

Macroscopically, a collision between two spherical nuclei is
fully characterized by the colliding energy and the
impact parameter. However, particle emission is the result of microscopic
processes (parton--parton or nucleon--nucleon collisions) which are not
described by these macroscopic variables. For fixed energy and impact
parameter,
this results in statistical fluctuations
of global macroscopic quantities such as the sphericity tensor $S^\bot_{ij}$.
Now, since particles are created independently at different points in the
system,
it is reasonable to assume that $S^\bot_{ij}$ is the sum of a large number of
independent random contributions. Then the central limit theorem states that
the probability law of $S^\bot_{ij}$ is gaussian.
To avoid having to deal with too many indices, we rearrange the three
independent components of $S^\bot_{ij}$ into a 3-vector $S$ whose components
$S_i$ are defined as
\beq
S_1&=&S^\bot_{11}+S^\bot_{22}={\cal E}\nonumber\\
S_2&=&S^\bot_{11}-S^\bot_{22}={\cal E}\alpha\,{\rm cos}\,2\theta\\
S_3&=&2S^\bot_{12}={\cal E}\alpha\,{\rm sin}\,2\theta .\nonumber
\eeq
The most general gaussian probability distribution for $S_i$ is of the form
\beq\label{centrlimit}
{{\rm d}P\over {\rm d}S_1{\rm d}S_2{\rm d}S_3}\propto
{\rm exp}\left[ -\left( {^tS}-{^t\bar S}\right)T^{-1}
\left( S-\bar S\right)/2 \right]
\eeq
where $\bar S=\langle S\rangle$ is the average value of $S$ and
$T$ is the $3\times 3$ covariance matrix defined by
\beq
T_{ij}=\langle S_iS_j\rangle -\bar S_i\bar S_j.
\eeq
If one considers two independent random variables $S$ and $S'$ with
gaussian probabilities such as (\ref{centrlimit}), with respective
covariance matrices $T$ and $T'$, the probability
of $S+S'$ is gaussian, with a covariance matrix equal to
$T+T'$. Thus $T_{ij}$ is proportional to the number of particles
$M$ used in the analysis.

Since the system is symmetric with respect to the reaction plane for
spherical nuclei, so must be the probability (\ref{centrlimit}).
We choose $x$ to be the direction of the impact parameter. Then the
symmetry with respect to the reaction plane changes $\theta$ into
$-\theta$. Eq.(\ref{centrlimit}) is invariant under this
transformation if
\beq
\bar S_3&=&0\nonumber\\
T_{13}&=&T_{23}=0.
\eeq
Thus $\bar S$ depends only on two parameters
\beq
\bar S_1&=&\bar{\cal E}\nonumber\\
\bar S_2&=&\bar{\cal E}\bar\alpha.
\eeq
$\bar\alpha$ represent the anisotropy associated with $\bar S$, that is the
anisotropy in the emission law. Although $\bar S$ is the average value of $S$,
$\bar\alpha$ is in general not equal to the average value of $\alpha$,
$\langle\alpha\rangle$~:
for an isotropic emission, $\bar\alpha=0$, but in general $\alpha>0$ with a
finite number of particles. While $\bar\alpha$ represents the anisotropy
associated with macroscopic effects, for instance with collective flow,
the average value $\langle\alpha\rangle$ also receives a contribution from
finite multiplicity
fluctuations.

If the emission is weakly anisotropic, i.e. if $\bar\alpha\ll 1$,
the covariance matrix $T$ is approximately
the same as for an isotropic distribution. For an isotropic distribution,
the probability (\ref{centrlimit}) must be invariant under rotations, that
is under transformations $\theta\rightarrow\theta+\theta_0$ with $\theta_0$
fixed. This implies
\beq
T_{12}&=&0\nonumber\\
T_{22}&=&T_{33}.
\eeq
Then the probability (\ref{centrlimit}) can be integrated over ${\cal E}$
with the result\cite{olli92}
\beq\label{genprob}
{{\rm d}P\over {\rm d}\alpha {\rm d}\theta}= {4\alpha\over\pi\sigma^2}
{\rm exp}\left( -\frac{\bar\alpha^2+\alpha^2-2\alpha\bar\alpha\,\cos\,2\theta}
{\sigma^2}\right)
\eeq
with
\beq
\sigma={\sqrt{2T_{22}}\over \bar{\cal E}}.
\eeq
$\sigma$ is the order of magnitude of the anisotropy created by statistical
fluctuations
alone, as can be seen easily if $\bar\alpha=0$: the probability distribution
for $\alpha$, apart from the preexponential factor $\alpha$ which arises from
the
jacobian transforming $(\alpha,{\cal E},\theta)$ into $(S_1,S_2,S_3)$,
is a gaussian of width $\sigma/\sqrt{2}$. Note that since both
$T_{22}$ and $\bar{\cal E}$ are proportional to the number of particles
$M$ used in the analysis, $\sigma$ scales like $1/\sqrt{M}$.
For an uncorrelated emission of identical particles, one finds
$\sigma=(1/\sqrt{M})\langle w^2\rangle/\langle w\rangle^2$, where
$w$ is the weight of the particle in the sphericity tensor, Eq.(\ref{sbot}).
We have chosen to normalize Eq.(\ref{genprob}) to unity in the interval
$[0,\pi/2]$ (rather than $[-\pi/2,\pi/2]$ because the function is even),
that is $\int_0^{\pi/2}d\theta\int_0^1d\alpha (dP/d\alpha d\theta)=1$.
This will be checked below.

It is convenient to express the probability as a function of the
scaled quantities
\beq\label{barchi}
\chi=\alpha/\sigma\ \ {\rm and}\ \ \bar\chi=\bar\alpha/\sigma
\eeq
Eq.(\ref{genprob}) becomes then
\beq\label{chigenprob}
{{\rm d}P\over {\rm d}\chi {\rm d}\theta}= {4\chi\over\pi}
{\rm exp}\left( -\bar\chi^2-\chi^2+2\bar\chi\chi\,\cos\,2\theta \right)
\eeq
This equation only involves the dimensionless parameter $\bar\chi$,
which will play a crucial role in our analysis. Physically, $\bar\chi$
represents the ratio of the anisotropy $\bar\alpha$ generated by dynamical
collective effects to the typical anisotropy $\sigma$ yielded by statistical
fluctuations. Note that strictly speaking, Eq.(\ref{chigenprob})
also involves the parameter $\sigma$ since $\chi$ varies from 0 to $1/\sigma$.
However, $\sigma\ll 1$ for a large system and the probability
(\ref{chigenprob})
decreases exponentially for $\chi\gg 1$. Thus we will let $\chi$ vary
from 0 to $+\infty$ when we integrate over $\chi$.

Eq.(\ref{chigenprob}) can be integrated over $\theta$ using the modified
Bessel function $I_0$ defined in Eq.(\ref{defi0}). One gets then the
distribution of the scaled anisotropy $\chi$:
\beq\label{chiprob}
{{\rm d}P\over {\rm d}\chi}= 2\chi \exp(-\bar\chi^2-\chi^2)I_0(2\bar\chi\chi).
\eeq
To check that this distribution is normalized to unity, we
integrate (\ref{chiprob}) by parts and use the relation $I_1(z)=dI_0/dz$. Then
the formula
\beq\label{normbessel}
\int_0^{+\infty}{\exp(-\chi^2)I_\nu (2\bar\chi\chi)d\chi}={\sqrt{\pi}\over 2}
\exp(\bar\chi^2/2)I_{\nu/2}(\bar\chi^2/2)
\eeq
with $\nu=1$ gives the result, using the fact that
$I_{1/2}(z)=2\sinh z/\sqrt{2\pi z}$.

In principle, the distribution (\ref{chiprob}) can be compared to
the experimental distribution of the anisotropy $\alpha$:
$\bar\chi$ and the scale factor $\sigma$ can be fitted so as
to obtain the best agreement with the data.
However, this may be difficult in practise, as we are going to see shortly.
Fig.1 displays $(1/\chi)(dP/d\chi)$ (that is the distribution
(\ref{chiprob}) divided by the factor $\chi$ arising from the
jacobian), for three values of the parameter $\bar\chi$.
For $\bar\chi=0$ (no collective flow),
this quantity would be a gaussian of width $1/\sqrt{2}$
centered at $\chi=0$ as can
be seen from Eq.(\ref{chiprob}). On the other hand,
if $\bar\chi > 1$, one easily shows that
$(1/\chi)(dP/d\chi)$ reaches its maximum at a non vanishing value
of $\chi$, which becomes closer to $\bar\chi$ as $\bar\chi$ increases.
This can be used\cite{olli92} as a signature of collective flow.
If $\bar\chi < 1$, however, the maximum is reached at $\chi=0$ and
the distribution is very close to a gaussian
as illustrated in Fig.1 for $\bar\chi=0.5$. Thus, the
effect of $\bar\chi$ is simply to increase the width of the
$\chi$-distribution, compared to $\bar\chi=0$.
Since the $\alpha$ distribution has the same shape as
for $\bar\chi=0$, it is impossible to extract $\bar\chi$
from the anisotropy distribution alone.

Eq.(\ref{chigenprob}) can also be integrated
over $\chi$, which yields the probability distribution of the angle $\theta$
\beq\label{pteta}
{{\rm d}P\over {\rm d}\theta}=
{2\over\pi}\exp(-\bar\chi^2)\left\{ 1+\sqrt{\pi}\bar\chi\cos\,2\theta
\left[ 1+{\rm erf}(\bar\chi\cos\,2\theta)\right]
\exp\left(\bar\chi^2\cos^2\,2\theta\right) \right\}
\eeq
where
\beq
{\rm erf}(x)={2\over\sqrt{\pi}}\int_0^x{{\rm e}^{-t^2}{\rm d}t}
\eeq
is the standard error function. ${\rm d}P/{\rm d}\theta$ is a
decreasing function of $\theta$.
If statistical fluctuations are large compared to the
dynamical anisotropy, that is if $\bar\chi\ll 1$,
Eq.(\ref{pteta}) reduces to
\beq\label{smallchi}
{{\rm d}P\over {\rm d}\theta}=
{2\over\pi}\left(1+\sqrt{\pi}\bar\chi\cos\,2\theta\right)+{\cal O}(\bar\chi^2).
\eeq
The anisotropy results here in a small deviation, with amplitude
proportional to $\bar\chi$,
from the constant value corresponding to isotropic emission.
On the other hand, in the limit where $\bar\chi\gg 1$ (strong anisotropy),
Eq.(\ref{pteta}) becomes
\beq\label{bigchi}
{{\rm d}P\over {\rm d}\theta}={4\bar\chi\over\sqrt{\pi}}
\exp\left(-4\bar\chi^2\theta^2\right).
\eeq
In this case, the probability is a gaussian
of width $1/(2\sqrt{2}\bar\chi)\ll 1$, centered at $\theta=0$.
Note that Eq.(\ref{pteta}) is of little practical use~:
$\theta$ is measured from the $x$--axis which we have chosen,
in this section, to be the direction of impact parameter $x$.
But since there is no direct access to this direction in the
experiment, $\theta$, unlike $\alpha$, is not
an observable. We do not get any physical information from $\theta$
unless it is correlated with another independent evaluation of the
reaction plane.

\section{Reaction plane correlations}

If one measures the reaction plane from $S^\bot$ in two
separate rapidity intervals with the same multiplicity,
one obtains two angles $\theta_1$ and $\theta_2$
(measured from an arbitrary fixed direction) which are two
measurements of the reaction plane. The determination is
reliable only if $\theta_1$ and $\theta_2$ are strongly correlated.
In this section, we calculate the probability distribution
${{\rm d}P_{\rm corr}/{\rm d}\theta}$
of the relative angle
$\theta\equiv\theta_1-\theta_2$. If this probability is
flat, $\theta_1$ and $\theta_2$ are uncorrelated and no
conclusion can be drawn concerning the occurrence of collective
flow. We expect this to be the case if anisotropy is small
or statistical fluctuations are large, that is if $\bar\chi\ll 1$.
If, on the other hand, $\bar\chi\gg 1$, we expect
${{\rm d}P_{\rm corr}/{\rm d}\theta}$
to be strongly peaked at $\theta=0$.

We shall assume that the rapidity intervals are well
separated, so that there is no correlation between them,
and consider the two corresponding sphericity tensors as statistically
independent. Then $\theta_1$ and $\theta_2$ are two independent random
variables. We further assume that macroscopic quantities (fluid
velocity, energy density) are invariant under Lorentz boosts along
the collision axis\cite{bjorken} and postpone the discussion
of this point to section 6. Since the sphericity tensor $S^\bot$
involves transverse coordinates only, it is also boost invariant. Thus
$\bar\alpha$ and $\sigma$ are the same for both rapidity intervals and
$\theta_1$ and $\theta_2$ have the same probability distribution,
given by Eq.(\ref{pteta}). The correlation function is then given by
\beq
\label{pcordef}
{{\rm d}P_{\rm corr}\over{\rm d}\theta}
&=&{1\over 2}\int_{-\pi/2}^{\pi/2} {{\rm d}\theta_1
\int_{-\pi/2}^{\pi/2} {{\rm d}\theta_2\,
{{\rm d}P\over{\rm d}\theta}(\theta_1)\,{{\rm d}P\over{\rm d}\theta}(\theta_2)
\,\delta(\theta-\theta_1-\theta_2)}}\nonumber\\
&=& {1\over 2}\int_{-\pi/2}^{\pi/2} {{\rm d}\theta_1\,
{{\rm d}P\over{\rm d}\theta}(\theta_1)\,{{\rm d}P\over{\rm d}\theta}
(\theta_1-\theta)}
\eeq
where ${\rm d}P/{\rm d}\theta$ is given by Eq.(\ref{pteta}).
The factor $1/2$ normalizes ${\rm d}P_{\rm corr}/{\rm d}\theta$
to unity between $0$ and $\pi/2$.

The integration can be carried out analytically (see Appendix A), which
yields the result
\beq\label{correl}
{{\rm d}P_{\rm corr}\over{\rm d}\theta}&=&
{\rm e}^{-\bar\chi^2}\left\{
{2\over\pi}(1+\bar\chi^2)+\bar\chi^2\left[
\cos\,2\theta\, (I_0+{\bf
L}_0)(\bar\chi^2\cos\,2\theta)\right.\right.\nonumber\\
& & \mbox{}\left.\left.+(I_1+{\bf L}_1)(\bar\chi^2\cos\,2\theta)\right]\right\}
\eeq
where $I_0$ and $I_1$ are modified Bessel functions of the first kind and
${\bf L}_0$ and ${\bf L}_1$ are modified Struve functions.
Eq.(\ref{correl}) reduces to
\beq\label{smallchibis}
{{\rm d}P_{\rm corr}\over{\rm d}\theta}={2\over\pi}+\bar\chi^2\cos\,2\theta
\eeq
if $\bar\chi\ll 1$, and to
\beq\label{bigchibis}
{{\rm d}P_{\rm corr}\over{\rm d}\theta}=\sqrt{8\over\pi}\,\bar\chi
\exp(-2\bar\chi^2\theta^2)
\eeq
if $\bar\chi\gg 1$. These asymptotic forms can be deduced directly from
Eqs.(\ref{smallchi}), (\ref{bigchi}) and (\ref{pcordef}).
Note that for small $\bar\chi$, the correlations (deviations from
a flat probability) are of order $\bar\chi^2$. Thus, when
statistical fluctuations become larger than dynamical
effects, correlations decrease very quickly.
On the other hand, if $\bar\chi\gg 1$,
${{\rm d}P_{\rm corr}/{\rm d}\theta}$
is the convolution of two identical gaussians (\ref{bigchi}),
that is a gaussian of width $1/(2\bar\chi)$.
The correlation function given by Eq.(\ref{correl}) is
displayed in Fig.2 for three values of $\bar\chi$, together with
the approximations (\ref{smallchibis}) and (\ref{bigchibis}). One
sees that these approximations are very good for
$\bar\chi\le 0.5$ and $\bar\chi\ge 2$ respectively.

A measure of the correlation strength is obtained by forming the
ratio of the number of events with $\theta>45^\circ$
to the number of events with $\theta<45^\circ$. This ratio
is equal to 1 if there is no correlation between reaction planes
and vanishes if they are strongly correlated. Integrating
Eq.(\ref{correl}) over $\theta$, one obtains a simple analytic expression
for this ratio (see Appendix A):
\beq\label{ratio}
{N_{\theta>45^\circ}\over N_{\theta<45^\circ}}
={\displaystyle\int_{\pi/4}^{\pi/2}{{{\rm d}P_{\rm corr}\over{\rm d}\theta}{\rm
d}\theta}\over
\displaystyle\int_0^{\pi/4}{{{\rm d}P_{\rm corr}\over{\rm d}\theta}{\rm
d}\theta}}
={1\over 2\exp(\bar\chi^2)-1}.
\eeq
This quantity is displayed in Fig.3 as a function of $\bar\chi$. As expected,
it decreases from 1 to 0 when $\bar\chi$ goes from 0 to $+\infty$. One
observes that this ratio differs significantly from unity already
for modest values of $\bar\chi$, so that collective effects
should be seen easily by measuring reaction plane correlations.
Any deviation of $dP_{\rm corr}/d\theta$
from a constant value can be attributed to collective flow
(within our hypotheses), which makes this signature less ambiguous
than that associated with the distribution of anisotropy (see Fig.1 and the
corresponding discussion in section 3).

The ratio in Eq.(\ref{ratio}) can be measured directly, and from its
value one deduces the value of $\bar\chi$. One may then check
whether Eq.(\ref{correl}) reproduces the observed behavior of
the correlation function.

\section{Relation to anisotropy}

The correlation between reaction planes clearly does not exhaust all
the information we get from the sphericity tensor analysis.
We also have a measurement of the anisotropy $\alpha$.
The value of $\bar\chi$ one gets from the analysis of
plane correlations fixes
the shape of the $\chi$ distribution, using Eq.(\ref{chiprob}). In order to get
the $\alpha$ distribution, the scale factor $\sigma$ is required
(see Eq.(\ref{barchi})). As we shall see shortly,
this quantity can be determined through the measured average value
of $\alpha$, which we denote by $\langle\alpha\rangle$.
It is directly proportional to $\sigma$.

When we studied reaction plane correlations in section 4,
only half of the detected particles (at most) could
be used to construct the sphericity tensor since we needed two
independent evaluations of the reaction plane in each event.
On the other hand, when measuring the anisotropy distribution,
it is better to use all the particles detected in the
central rapidity region, in order to
minimize statistical fluctuations.
Then the value of $\bar\chi$ which has been determined
from reaction plane correlation cannot be used directly in
analyzing the $\alpha$-distribution. Since $\sigma$ scales
like $1/\sqrt{M}$ and $\bar\alpha$ is independent of
$M$, $\bar\chi$ must be scaled like $\sqrt{M}$.
If, for instance, the set of measured particles is
divided into two approximately equal subsets for the
measurement of reaction plane correlations, and then used
as a whole for measuring the anisotropy distribution,
the value of $\bar\chi$ must be multiplied by $\sqrt{2}$.

Let us now calculate the average value of $\alpha$:
\beq\label{alphamoy}
\langle\alpha\rangle=\sigma\langle\chi\rangle
=\sigma{\int{\chi(dP/d\chi)d\chi}\over\int{(dP/d\chi)d\chi}}
\eeq
The denominator of this expression is equal to unity since
$dP/d\chi$ given by Eq.(\ref{chiprob}) is normalized to unity.
To calculate the numerator, we integrate Eq.(\ref{chiprob})
by parts and use the relation $dI_1/dz=(I_0(z)+I_2(z))/2$
and Eq.(\ref{normbessel}) to calculate the remaining integrals.
The result is
\beq\label{chimoy}
\langle\chi\rangle=
{\sqrt{\pi}\over 2}\left[
(1+\bar\chi^2)I_0(\bar\chi^2/2)+\bar\chi^2I_1(\bar\chi^2/2)\right]
\exp(-\bar\chi^2/2).
\eeq
Thus, if one measures $\bar\chi$ and $\langle\alpha\rangle$, the last
two equations gives the scale factor
$\sigma$ and thus $\bar\alpha=\sigma\bar\chi$.
{}From Eq.(\ref{alphamoy}) one gets immediately
\beq
{\bar\alpha\over\langle\alpha\rangle}={\bar\chi\over\langle\chi\rangle}.
\eeq
This quantity is displayed in Fig.4. For small anisotropy,
$\bar\chi\ll 1$, the $\chi$ distribution (\ref{chiprob}) is
gaussian and one gets
\beq\label{smallchiter}
{\bar\alpha\over\langle\alpha\rangle}\simeq {2\over\sqrt{\pi}}\bar\chi.
\eeq
In this case most of the observed anisotropy results from
fluctuations and the dynamical component $\bar\alpha$ is only
a small fraction of the average anisotropy. On the other hand,
for $\bar\chi\gg 1$, statistical fluctuations become negligible
and the average anisotropy $\langle\alpha\rangle$ is close to
$\bar\alpha$. Asymptotically,
\beq\label{bigchiter}
{\bar\alpha\over\langle\alpha\rangle}\simeq 1-{1\over 4\bar\chi^2}.
\eeq
Once $\bar\alpha$ and $\sigma$ are determined, the measured
$\alpha$ distribution can be compared to the theoretical
prediction, Eq.(\ref{chiprob}).

\section{Discussion}

Let us recall and discuss the hypotheses on which
our calculations are based. The first hypothesis was made
at the beginning of section 3, where we assumed that the
sphericity tensor $S_{ij}$ has a gaussian distribution.
This is true if it can be considered as the sum of a large
number of independent sources, which is a reasonable assumption
if the nucleon--nucleon collisions
creating the particles are incoherent. However, deviations
from this behavior can occur due to jets, which
result in strongly correlated, strongly anisotropic emission
of particles. At very high energies, for instance at LHC,
one expects a large number of jets per event, and pairs of
jets can be considered independent so that our statement
holds~: the only consequence is that the number of independent
sources is the number of jets rather than the number of
produced particles, so that we expect larger statistical
fluctuations. On the other hand, if only a few jets (say, one or two)
are produced in each event, which may be the case at lower energies,
significant deviations from the gaussian distribution may
occur. These would cause deviations of the anisotropy
and angle distributions from the shapes predicted by our
model, given by Eqs.(\ref{chiprob}) and (\ref{correl}).

The second hypothesis was made in section 4 where we treated
the sphericity tensors measured in two separate rapidity intervals
as independent variables. This is not strictly
true if the rapidity intervals are too close to each other~:
for instance, a resonance decay or a pair of jets can give contributions to
both rapidity intervals.
This can be avoided by taking two remote rapidity intervals. This
will be possible at RHIC and LHC if detectors have a large
rapidity acceptance. In current experiments at CERN and
Brookhaven, the rapidity window is not so large and one may
be constrained to work with adjacent rapidity intervals.
One may check directly, as a test of statistical independence,
that the value of $\bar\chi$ deduced from the analysis indeed
scales like $\sqrt{M}$, with $M$ the number of particles used
in the analysis.

We also assumed in section 4 that the probability distribution
of $S^\bot$ was the same for both rapidity intervals as a consequence
of Bjorken's scenario\cite{bjorken} for the fluid evolution. This
scenario is known to be unrealistic in current experiments since
the measured rapidity distributions are not flat. However,
this hypothesis is not crucial here. It simplified the calculations
since we used the same value of the scaled anisotropy $\bar\chi$
for both rapidity intervals. Without this hypothesis, we wouldn't
have obtained analytical expressions, but the qualitative ideas
would remain the same. In particular, the correlation between reaction
planes could still be used to identify collective flow.
Note further that recourse to Bjorken's scenario can be avoided
for a symmetric collision:
if the two rapidity intervals are chosen symmetric of each other
in the center of mass frame, they are equivalent by symmetry and the
value of $\bar\chi$ is the same for both. Even more generally,
it is reasonable to assume that the anisotropy $\bar\alpha$ depends
only weakly on the rapidity since it is not much affected by
the longitudinal expansion\cite{olli92}. If the two rapidity intervals
have the same multiplicity, the statistical fluctuations $\sigma$
should also be comparable and therefore there is no reason for
$\bar\chi$ to change drastically from one rapidity interval to another.

Although one must be careful in applying our results according
to the above discussion, reaction plane correlations appear to provide
a powerful tool for identifying collective flow. The ratio defined in
Eq.(\ref{ratio}) is very sensitive to collective effects:
for $\bar\chi=0.3$, this ratio is about 0.85, and such a deviation
from unity would be seen clearly in an experiment.
We have shown\cite{olli92} that the anisotropy distribution
alone should allow to identify collective flow at AGS and SPS
with heavy nuclei (Au or Pb projectiles). With the plane correlation method
presented here,
the possibility is not excluded that collective effects could
be seen with lighter projectiles, for instance with $^{32}$S or
$^{40}$Ca beams on heavy targets.

\section*{Acknowledgements}

I thank H. Gutbrod for stimulating discussions and J.-P. Blaizot
for careful reading of the manuscript.

\renewcommand{\theequation}{A.\arabic{equation}}
\setcounter{equation}{0}
\section*{Appendix A: Derivation of the correlation function}

Inserting Eq.(\ref{chigenprob}) in Eq.(\ref{pcordef}) one may write
the correlation function in the form
\beq\label{chi12}
{{\rm d}P_{\rm corr}\over{\rm d}\theta}&=&
{8\over\pi^2}\exp(-2\bar\chi^2)
\int_0^{+\infty}\chi_1 d\chi_1\int_0^{+\infty}\chi_2 d\chi_2
\exp\left(-\chi_1^2-\chi_2^2\right) \\
&&\times\int_{-\pi/2}^{\pi/2}{d\theta_1
\exp\left[
2\bar\chi\chi_1\cos\,2\theta_1
+2\bar\chi\chi_2\cos\,2(\theta_1-\theta)\right]
}\nonumber
\eeq
This can be integrated over $\theta_1$ by using the modified
Bessel function $I_0$:
\beq\label{defi0}
\int_{-\pi/2}^{\pi/2}{d\theta_1\exp\left(A\,\cos\,2\theta_1+B\,\sin\,2\theta_1\right)}
=\pi I_0\left(\sqrt{A^2+B^2}\right)
\eeq
This gives
\beq\label{rphi}
{{\rm d}P_{\rm corr}\over{\rm d}\theta}&=&
{8\over\pi}\exp(-2\bar\chi^2)
\int_0^{+\infty}\chi_1 d\chi_1\int_0^{+\infty}\chi_2 d\chi_2
\exp\left(-\chi_1^2-\chi_2^2\right) \nonumber\\
&&\mbox{}\times
I_0\left( 2\bar\chi\sqrt{\chi_1^2+\chi_2^2+2\chi_1\chi_2\,\cos\,2\theta}\right)
\eeq
Introducing polar coordinates in the $(\chi_1,\chi_2)$ plane,
defined by $\chi_1=r\,\cos\phi$, $\chi_2=r\,\sin\phi$, with $r\ge 0$ and
$0\le\phi\le\pi/2$, the integral over
$r$ is of the type
\beq
\int_0^{+\infty}{dr\, r^3\exp(-r^2)I_0(ar)}={1\over 2}\left(1+{a^2\over
4}\right)
\exp\left({a^2\over 4}\right)
\eeq
as may be checked by expanding $I_0$.
Eq.(\ref{rphi}) thus becomes
\beq\label{phi}
{{\rm d}P_{\rm corr}\over{\rm d}\theta}&=&
{2\over\pi}e^{-\bar\chi^2}
\int_0^{\pi/2}{ d\phi\,\sin 2\phi
\left[ 1+\bar\chi^2\left( 1+\sin 2\phi\cos\,2\theta\right)\right]}\nonumber\\
&&\mbox{}\times\exp\left(\bar\chi^2\sin 2\phi\cos\,2\theta\right)
\eeq
Note that the integrand is invariant under $\phi\rightarrow\pi/2-\phi$,
which reflects the fact that $\chi_1$ and $\chi_2$ play symmetric roles
in Eq.(\ref{chi12}). The integration range can then be restricted to
the interval $[0,\pi/4]$. Making the change of variables
$\pi/2-2\phi\rightarrow\phi$, the integral can be expressed in
terms of the modified Bessel functions
\beq\label{bessel}
I_0(z)&=&{2\over\pi}\int_0^{\pi/2}{\cosh(z\cos\phi)d\phi}\nonumber\\
I_1(z)&=&{2z\over\pi}\int_0^{\pi/2}{\sin^2\phi\cosh(z\cos\phi)d\phi}
\eeq
and the modified Struve functions\cite{bateman} ${\bf L}_0(z)$ and ${\bf
L}_1(z)$
which have expressions similar to $I_0$ and $I_1$, with $\cosh$ replaced
by $\sinh$. After an integration by parts, Eq.(\ref{phi}) yields the
result
\beq\label{correl1}
{{\rm d}P_{\rm corr}\over{\rm d}\theta}&=&
{\rm e}^{-\bar\chi^2}\left\{
{2\over\pi}(1+\bar\chi^2)+\bar\chi^2\left[
\cos\,2\theta\, (I_0+{\bf
L}_0)(\bar\chi^2\cos\,2\theta)\right.\right.\nonumber\\
&&\mbox{}\left.\left.+(I_1+{\bf L}_1)(\bar\chi^2\cos\,2\theta)\right]\right\}
\eeq
identical to Eq.(\ref{correl}).
Let us check that this expression is normalized to unity
when integrated from $0$ to $\pi/2$.
the functions $\cos 2\theta I_0(\bar\chi^2\cos 2\theta)$ and
$I_1(\bar\chi^2\cos 2\theta)$ are odd in $\cos 2\theta$ and thus
do not contribute to the integral. Using the relations
$I_1(z)=dI_0/dz$ and ${\bf L}_1(z)=d{\bf L}_0/dz+2/\pi$ and
the definitions (\ref{bessel}), the
integrals of Struve functions can be expressed in terms of
integrals of Bessel functions, which can be found in the
literature or calculated directly by power series expansion:
\beq\label{lemme1}
\int_0^{\pi/2}{d\theta\,{\bf L}_1(\bar\chi^2\cos\theta)}
&=&-1+\int_0^{\pi/2}{d\phi \cos\phi I_0(\bar\chi^2\cos\phi)}
\nonumber\\
&=&-1+{\sinh\bar\chi^2\over\bar\chi^2}
\eeq
and
\beq\label{lemme2}
\int_0^{\pi/2}{d\theta \cos\theta\,{\bf L}_0(\bar\chi^2\cos\theta)}
&=&\int_0^{\pi/2}{d\phi \cos\phi I_1(\bar\chi^2\cos\phi)}
\nonumber\\
&=&{\cosh\bar\chi^2-1\over\bar\chi^2}.
\eeq
Normalization of the probability (\ref{correl1}) follows
immediately.

Let us finally calculate the ratio defined in Eq.(\ref{ratio}).
Since the correlation function (\ref{correl1}) is normalized
to unity, we only need to calculate
$\int_{\pi/4}^{\pi/2}{({\rm d}P_{\rm corr}/{\rm d}\theta)d\theta}$.
On this interval, all the terms in Eq.(\ref{correl1}) give a
non vanishing contribution. However, using Eqs.(\ref{lemme1}) and
(\ref{lemme2}) one obtains
\beq
\int_{\pi/4}^{\pi/2}{d\theta\,\cos 2\theta\,I_0(\bar\chi^2\cos 2\theta)}
=-{1\over 2}-\int_{\pi/4}^{\pi/2}{d\theta\,{\bf L}_1(\bar\chi^2\cos 2\theta)}
\eeq
and
\beq
\int_{\pi/4}^{\pi/2}{d\theta\,I_1(\bar\chi^2\cos 2\theta)}
=-\int_{\pi/4}^{\pi/2}{d\theta\,\cos 2\theta\,{\bf L}_1(\bar\chi^2\cos
2\theta)}
\eeq
The terms involving Bessel functions and Struve functions in Eq.(\ref{correl})
thus cancel pairwise in the integration and one gets
\beq
\int_{\pi/4}^{\pi/2}{{{\rm d}P_{\rm corr}\over{\rm d}\theta}d\theta}
={1\over 2}\exp(-\bar\chi^2),
\eeq
from which Eq.(\ref{ratio}) follows immediately.

\section*{Figure captions}

\noindent{\bf Fig. 1~:} The solid lines display the values of
$(1/\chi)(dP/d\chi)$, given by Eq.(\ref{chiprob}), as a function
of $\chi$ for three values of the parameter $\bar\chi$.
The dashed line is an approximation to the curve
$\bar\chi=0.5$ by a gaussian corresponding
to the same average value $\langle\chi\rangle$ of $\chi$, calculated
from Eq.(\ref{chimoy}) (see section 5). All curves are normalized to
unity~: $\int_0^{+\infty}(1/\chi)(dP/d\chi)d\chi=1$.

\noindent{\bf Fig. 2~:} Solid lines: correlation function defined by
Eq.(\ref{correl}) for three values of $\bar\chi$.
Dashed line: Small $\bar\chi$ approximation, Eq.(\ref{smallchibis}).
Dot-dashed line: large $\bar\chi$ approximation, Eq.(\ref{bigchibis}).

\noindent{\bf Fig. 3~:} Ratio defined in Eq.(\ref{ratio}) as a function
of the dimensionless parameter $\bar\chi$.

\noindent{\bf Fig. 4~:} Solid line: ratio of the ``dynamical'' anisotropy
$\bar\alpha$ to the measured average value of $\alpha$, denoted by
$\langle\alpha\rangle$, as a function of $\bar\chi$.
Dashed line:
small $\bar\chi$ approximation, Eq.(\ref{smallchiter}). Dot-dashed line:
large $\bar\chi$ approximation, Eq.(\ref{bigchiter}).
\newpage
\centerline{\ \ }
\includegraphics{fig1.ps}
\newpage
\centerline{\ \ }
\includegraphics{fig2.ps}
\newpage
\centerline{\ \ }
\includegraphics{fig3.ps}
\newpage
\centerline{\ \ }
\includegraphics{fig4.ps}
\end{document}